\documentclass[aps,showpacs,amsmath,preprint,amssymb,12pt]{revtex4}

\usepackage{graphicx}

\usepackage{bm}

\begin{document}
{~}
\vspace*{1cm}

\title{Rotating Kaluza-Klein Multi-Black Holes with G\"odel Parameter 
\vspace{1cm}
}
\author{${}^{1}$Ken Matsuno\footnote{E-mail: matsuno@sci.osaka-cu.ac.jp}, 
  ${}^{1}$Hideki Ishihara\footnote{E-mail: ishihara@sci.osaka-cu.ac.jp}, 
  ${}^{1}$Toshiharu Nakagawa\footnote{E-mail: tosiharu@sci.osaka-cu.ac.jp} 
  and 
 ${}^{2}$ Shinya Tomizawa\footnote{E-mail: tomizawa@post.kek.jp}}
\affiliation{ 
${}^{1}$Department of Mathematics and Physics,
Graduate School of Science, Osaka City University,
3-3-138 Sugimoto, Sumiyoshi-ku, Osaka 558-8585, Japan\\
${}^{2}$Cosmophysics Group, Institute of Particle and Nuclear Studies, KEK, Tsukuba, Ibaraki, 305-0801, Japan  
\vspace{2cm}
}

\begin{abstract}
We obtain new five-dimensional supersymmetric rotating multi-Kaluza-Klein black hole solutions  
with the G\"odel parameter  
in the Einstein-Maxwell system with a Chern-Simons term. 
These solutions have no closed timelike curve outside the black hole horizons. 
At the infinity, the space-time is effectively four-dimensional. Each horizon admits various lens space topologies $L(n;1)={\rm S^3}/{\mathbb Z}_n$ in addition to a round $\rm S^3$. The space-time can have outer ergoregions  
disjointed  
from the black hole horizons, as well as inner ergoregions   
attached to each horizon. 
We discuss the rich structures of ergoregions.
\end{abstract}

\preprint{OCU-PHYS 301 \ AP-GR 60}

\preprint{KEK-TH-1255}
\date{\today}
\maketitle

\section{Introduction}
In recent years, Kaluza-Klein black hole solutions have been studied by many 
authors in the context of string theory. Since Kaluza-Klein black hole solutions have compactified extra dimensions, 
the space-time effectively behaves as four-dimensions at the infinity.   
The first Kaluza-Klein black hole solutions with an extra twisted S$^1$ 
were found by Dobiasch and  Maison~\cite{DM} as vacuum solutions to the five-dimensional Einstein equation, 
and the features of the black hole were investigated by Gibbons and Wiltshire~\cite{GW}.  
The static charged Kaluza-Klein black holes were also found in the five-dimensional Einstein-Maxwell theory \cite{IM} and were generalized to the rotating case as the solutions in the  five-dimensional Einstein-Maxwell theory with a Chern-Simons term~\cite{NIMT}.  Supersymmetric rotating Kaluza-Klein black hole solutions were found by 
Gaiotto {\it et al}~\cite{GSY} and Elvang {\it et al}~\cite{EEMH}. 
Supersymmetric static multi-Kaluza-Klein black hole solutions were also constructed~\cite{IKMT}.   
These solutions were constructed on the self-dual Euclidean Taub-NUT space
in the framework of Gauntlett {\it et al}'s classification of the five-dimensional 
supersymmetric solutions~\cite{G}.   
In these solutions, at the infinity, the space-times asymptote to a twisted $\rm S^1$ bundle over the four-dimensional Minkowski space-time.   
Exact Kaluza-Klein black hole solutions which asymptote to the direct product of the four-dimensional Minkowski space-time and an $\rm S^1$ were also constructed~\cite{Myers,TIM,IMT}.

The squashing transformation\cite{IM, NIMT, Wang} 
is a useful tool to generate Kaluza-Klein 
black hole solutions 
from the class of {\it cohomogeneity-one} black hole solutions 
with the asymptotically flatness. 
Actually, Wang~\cite{Wang} regenerated the five-dimensional Kaluza-Klein 
black hole solution found by Dobiasch and Maison~\cite{DM} from the 
five-dimensional Myers-Perry black hole solution with two equal angular 
momenta~\cite{MP}. 
Applying the squashing transformation to the charged rotating black hole 
solutions with two equal angular momenta~\cite{CLP} 
in the five-dimensional Einstein-Maxwell theory with a Chern-Simons term, 
the present authors obtained the new Kaluza-Klein black hole 
solution~\cite{NIMT} in the same theory. 
This is the generalization of the Kaluza-Klein black hole solutions in Ref.~\cite{DM,GW,IM} and it describes a   
non-supersymmetric (non-BPS) black hole boosted in the direction of the extra dimension. 
One of the interesting features of the solution is that the horizon admits a prolate shape in addition to a round $\rm S^3$ by the effect of the rotation of black hole.

In the previous work~\cite{TIMN}, 
applying this squashing transformation to non-asymptotically flat Kerr-G\"odel black hole 
solutions~\cite{Gimon-Hashimoto}, 
we also constructed a new type of rotating Kaluza-Klein black hole solutions 
to the five-dimensional Einstein-Maxwell theory with a Chern-Simons term. 
Though the Kerr-G\"odel black hole solutions have closed timelike curves in the region away from the black hole, 
the squashed Kerr-G\"odel black hole solutions have no closed timelike curve outside the black hole horizons. 
In addition, the solution has two kinds of rotation parameters in the same direction of the extra dimension. 
These two independent parameters are associated with the rotations of the black hole and the universe. 
In the absence of a black hole, 
the solution describes the Gross-Perry-Sorkin (GPS) monopole 
which is boosted in the direction of an extra dimension 
and has an ergoregion by the effect of the rotation of the universe.

In this paper,   
taking a limit of parameters in the charged version of squashed 
Kerr-G\"odel black hole solutions introduced 
in the Appendix of ref~\cite{TIMN}, 
we construct new supersymmetric rotating Kaluza-Klein black hole solutions to 
the five-dimensional Einstein-Maxwell theory with a Chern-Simons term. 
These can be regarded as solutions generated by the squashing transformation   
of   
the supersymmetric Kerr-Newman-G\"odel black hole solutions~\cite{Herdeiro}.  
Like the squashed Kerr-G\"odel black holes in~\cite{TIMN}, 
these Kaluza-Klein black hole solutions have no closed timelike curve outside the black hole horizons. 
The space-time is asymptotically locally flat, i.e.,  at the infinity, 
the space-time approaches a twisted $\rm S^1$ fiber bundle over a four-dimensional Minkowski space-time. 
The horizons are a round $\rm S^3$, unlike known supersymmetric rotating Kaluza-Klein black hole solutions, 
where they are the squashed $\rm S^3$. 
We also generalize these solutions to multi-black hole solutions. 
In particular, we study two-black hole solutions. 
As will be shown later, each horizon admits various lens space topologies $L(n;1)={\rm S^3}/{\mathbb Z}_n$ 
($n$ : natural numbers) in addition to an ${\rm S^3}$ and ergoregions have rich structures.

The rest of this paper is organized as follows. First, following the results of classification of solutions of the five-dimensional minimal supergravity~\cite{G}, in section \ref{sec:solution}, we construct general solutions on the Gibbons-Hawing space, which is the Taub-NUT space in the special case. In section \ref{sec:single}, we present a new supersymmetric single black hole solution on the Taub-NUT space. 
In section \ref{sec:multi}, we study the multi-black hole solutions, in particular two black holes case.  We conclude our article with a discussion in section \ref{sec:summary}.

\section{Solutions}\label{sec:solution}

We consider the five-dimensional Einstein-Maxwell system with 
a Chern-Simons term.  
The action is given by  
\begin{eqnarray}
 S = \frac{1}{16 \pi G_5} \int d^5 x \sqrt{-g} \left[ 
   R - F_{\mu \nu } F^{\mu \nu } 
   - \frac{2}{3 \sqrt 3} \left( \sqrt{-g} \right)^{-1} 
   \epsilon ^{\mu \nu \rho \sigma \lambda } A_\mu F_{\nu \rho } F_{\sigma \lambda } 
   \right],                                            \label{action}
\end{eqnarray}
where $R$ is the five-dimensional scalar curvature, 
$\bm F = d \bm A$ is the 2-form of the five-dimensional gauge field 
associated with the gauge potential 1-form $\bm A$  
and $G_5$ is the five-dimensional Newton constant. 
Varying the action (\ref{action}), 
we can derive the Einstein equation
\begin{eqnarray}
 R_{\mu \nu } -\frac{1}{2} R g_{\mu \nu } 
 = 2 \left( F_{\mu \lambda } F_\nu^{ ~ \lambda } 
  - \frac{1}{4} g_{\mu \nu } F_{\rho \sigma } F^{\rho \sigma } \right), \label{Eineq}
\end{eqnarray}
and the Maxwell equation 
\begin{eqnarray}
 F^{\mu \nu}_{~~~; \nu} + \frac{1}{2 \sqrt 3 \sqrt{-g}} 
   \epsilon ^{\mu \nu \rho \sigma \lambda } F_{\nu \rho } F_{\sigma \lambda } = 0. 
 \label{Maxeq}  
\end{eqnarray}

We construct rotating multi-black hole solutions satisfying 
the equations (\ref{Eineq}) and (\ref{Maxeq}). 
The forms of the metric and the gauge potential one-form are
\begin{eqnarray}
 ds^2 &=& - H^{-2} \left[ dt + \alpha V ^\beta \left( d\zeta +\bm \omega \right) \right] ^2 + H ds^2_{\rm GH},  \label{metric} \\
 \bm A &=&  \frac{\sqrt 3}{2} H^{-1} \left[ dt + \alpha V ^\beta \left( d\zeta +\bm \omega \right) \right],  \label{gauge}      
\end{eqnarray}  
where the function $H$ and the metric $ds^2_{\rm GH} $ are given by
\begin{eqnarray}             
 H &=& 1 + \sum _i \frac{M_i}{\left| \bm R - \bm R _i \right|} ,  \label{H} \\
 ds^2_{\rm GH} &=& V^{-1} ds_{{\mathbb E}^3}^2 
                  + V \left( d\zeta +\bm \omega \right) ^2 ,   \label{GHmetric}\\
                 V^{-1} &=& \epsilon + \sum _i \frac{N_i}{\left| \bm R - \bm R _i \right|}, \label{nut}
\end{eqnarray}
respectively,
where $ds_{{\mathbb E}^3}^2=dx^2+dy^2+dz^2 $ is a metric on the three-dimensional Euclid space, ${\mathbb E}^3$, and ${\bm R}=(x,y,z)$ denotes a position vector on ${\mathbb E}^3$. The function $V^{-1}$ is a harmonic function 
on ${\mathbb E}^3$ with point sources located at 
$\bm R=\bm R_i:=(x_i,y_i,z_i)$, where the Killing vector field $\partial_\zeta$ has fixed points in the base space. 
The one-form $\bm \omega$, which is determined by
\begin{eqnarray}
\bm \nabla \times {\bm \omega}=\bm \nabla V^{-1}, 
\end{eqnarray}
has the explicit form
\begin{eqnarray}                 
 \bm \omega &=& \sum_{i} 
   N_i ~ \frac{z-z_i}{\left| \bm{R}-\bm{R}_i \right|} ~ 
   \frac{(x-x_i) dy -(y-y_i) dx}{(x-x_i)^2+(y-y_i)^2}, \label{omega}
\end{eqnarray}
where $M_i,N_i$ and $\alpha$ are constants, $\beta = \pm 1$, 
and $\epsilon = 0, 1$. 
The base space (\ref{GHmetric}) with the equations (\ref{nut}) and (\ref{omega}) is often called 
the Gibbons-Hawking space. 
In particular, the Gibbons-Hawking space with $\epsilon = 1$, $N_1 \not =0$ and $N_i = 0$ $(i \geq 2)$  
is the self-dual Euclidean Taub-NUT space, 
and the space with $\epsilon = 0$, $N_1 \not =0$ and $N_i = 0$ $(i \geq 2)$ 
is the four-dimensional Euclid space.

The solutions \eqref{metric}-\eqref{omega} coincide with several known black hole solutions.
For example, multi-black hole solutions on the multi-centered-Taub-NUT space~\cite{IKMT} are obtained by 
$\beta = -1 ,~ \epsilon = 1$ and $\alpha=0$.  Multi-black hole solutions on the multi-centered Eguchi-Hanson spaces~\cite{IKMT2} 
are obtained by setting $\beta =-1 $, $\epsilon =0$ and $\alpha=0$.

Restricting the solutions \eqref{metric} to the case with $\beta = 1$  and $\epsilon = 1$, 
we can obtain new black hole solutions. 
To avoid the existence of singularities and closed timelike curves outside the black hole horizons, 
we choose the parameters such that 
\begin{eqnarray}
M_i>0, \quad N_i>0 , \quad 0 \le \alpha^2 < 1 .
\end{eqnarray}
See Appendix \ref{PR} about the detail discussion.

\section{Single Black Hole with G\"odel Rotation}\label{sec:single}

First, we study the case of a single black hole, i.e., 
the case of $M_1 = M$, $N_1 = N$ and $ M_i =  N_i = 0$ $(i\ge 2)$.

\subsection{Metric and gauge potential }

In the single-black hole case, 
the metric \eqref{metric} and the gauge potential one-form \eqref{gauge} are expressed in the form
\begin{eqnarray}
 ds^2 &=& - H^{-2} 
\left[ dt + \alpha V  
 \left( d\zeta + N \cos \theta d\phi \right) \right] ^2
  \notag \\
        &&+ H 
 \left[ V^{-1} \left( dR^2 + R^2 d\Omega _{S^2} ^2 \right) 
        + V 
 \left( d\zeta + N\cos \theta d\phi \right) ^2 \right] ,    \label{metric0} \\
\bm A &=&  \frac{\sqrt 3}{2} H ^{-1} 
  \left[ dt + \alpha  V 
\left( d\zeta + N\cos \theta d\phi \right) \right] ,   \label{gauge0}  
\end{eqnarray}
where the functions $H$ and $V^{-1}$ can be written as
\begin{eqnarray}
 H = 1 + \frac{M}{R}, \quad V^{-1} = 1 + \frac{N}{R} ,   
\end{eqnarray}
respectively. 
$d\Omega _{S^2} ^2 = d\theta ^2 + \sin ^2 \theta d \phi ^2$ denotes 
the metric of the unit two-sphere.    
The coordinates run the ranges of $-\infty <t <\infty,\ -M < R < \infty,\  
0\leq \theta \leq \pi,\ 0\leq \phi \leq 2\pi$ and $0\le \zeta \le 2\pi L$.  
From the requirements for the absence of naked singularities and closed timelike curves (CTCs) outside the black hole horizon, the parameters are restricted to the region
\begin{eqnarray}
M > 0, \quad N > 0, \quad 0 \le \alpha ^2 < 1 .  \label{parahanni}
\end{eqnarray}
The constant $N$ is related to the size of the compactified radius $L$ at the infinity by 
\begin{eqnarray}
N=\frac{L}{2}n,
\end{eqnarray}
where $n$ is a natural number.

When the parameter $\alpha $ vanishes, 
the metric \eqref{metric0} and the gauge potential one-form \eqref{gauge0} 
coincide with the extreme case of the static charged Kaluza-Klein black hole solution \cite{IM}.

It should be noted that this solution \eqref{metric0} with $n=1$  
coincides with a limiting solution of the squashed Kerr-Newman-G\"odel 
black hole solution \cite{TIMN} by putting the parameters as
\begin{eqnarray}
 m=-q , \quad a=-2jq ,   
\label{limit}
\end{eqnarray}
and by identifying the coordinates and parameters as
\begin{eqnarray}
 t = \frac{r_\infty ^2 - m}{r_\infty ^2} T, \quad   
 R = \frac{r_\infty }{2} \frac{r ^2 - m}{r_\infty ^2 -r^2} , \quad 
 M = \frac{m}{2 r_\infty}, \quad 
 N = \frac{r_\infty }{2}, \quad 
 \alpha = \frac{2j \left( r_\infty ^2 -m \right) ^2}{r_\infty ^3} .   
\end{eqnarray}
About the explicit form of the squashed Kerr-Newman-G\"odel black hole solution, readers should see Appendix \ref{KN} in this article.

\subsection{Asymptotic structure and asymptotic charge}

Introduce a new coordinate $\psi := 2 \zeta / L$ with the periodicity of $\Delta\psi =4\pi$.  
In the neighborhood of the infinity, $R=\infty$, the metric \eqref{metric0} behaves as    
\begin{eqnarray}
 ds^2 \simeq -d\tilde t ^2 + dR ^2 + R ^2 d\Omega _{S^2} ^2 
  + \frac{L^2}{4} \left( 1-\alpha ^2 \right) \left( \frac{d\tilde \psi}{n} + \cos \theta d\phi \right) ^2,    
\end{eqnarray}
where we introduced the following coordinates $(\tilde t,\tilde \psi)$
\begin{eqnarray} 
\tilde t = \frac{t}{ \sqrt{1-\alpha ^2} } , \quad 
\tilde \psi = \psi - \frac{\alpha t}{ N \left( 1-\alpha ^2 \right) },  
\end{eqnarray}
which are chosen so that they are in the rest frame at the infinity. 
The asymptotic structure of the solution \eqref{metric0} is an asymptotically locally flat, 
i.e., the metric asymptotes to
a twisted constant $S^1$ fiber bundle over the four-dimensional Minkowski space-time and the spatial infinity has the structure of an $\rm S^1$ bundle over an $\rm S^2$ such that it is the lens space $L(n;1)={\rm S^3}/{\mathbb Z}_n$.

The Komar mass associated with the timelike Killing vector field  
$\partial_{\tilde t} $ at the infinity, $M$, the charge at the infinity, $Q$, and 
the angular momenta associated with the spacelike Killing vector fields 
$\partial_\phi $ and $\partial_{\tilde \psi}$ at the infinity, 
$J_\phi $ and $J_{\tilde \psi }$,  
can be obtained as 
\begin{eqnarray}
M &=&
\frac{3 L \left[ \left( \alpha ^2 + 2 \right) M + \alpha ^2 N \right]}
{ 8 \pi \sqrt{ 1-\alpha ^2 } G_5 } \frac{{\cal A} _{S^3}}{n} , \label{sbhmass} \\
Q &=& \frac{\sqrt 3 L M }{ 2 \pi G_5 } \frac{{\cal A} _{S^3}}{n} , \label{sbhcharge} \\
J_{\phi} &=& 0 , \\
J_{\tilde \psi} &=&  - \frac{\alpha L ^2 \left( 3 M + \alpha ^2 N \right)}
{ 8 \pi G_5 } \frac{{\cal A} _{S^3}}{n} \label{sbhangmom},
\end{eqnarray}
where ${\cal A} _{S^3}$ denotes the area of a unit $\rm S^3$.

\subsection{Horizon}

A black hole horizon exists at the position of the source for 
the harmonic functions $H$ and $V^{-1}$, i.e.,  $R=0$. 
In the coordinate system $(t ,R ,\theta , \phi ,\zeta)$, 
the metric \eqref{metric0} diverges apparently at $R=0$. 
In order to remove this apparent divergence, 
we introduce a new coordinate $v$ such that 
\begin{eqnarray} 
 dv = dt - \sqrt{\left( 1 + \frac{M}{R} \right) ^3 \left( 1 + \frac{N}{R} \right) } dR . 
\end{eqnarray}

Then, near $R = 0$, the metric \eqref{metric0} behaves as 
\begin{eqnarray} 
 ds^2 \simeq -2 \sqrt{\frac{N}{M}} dvdR 
+ MN \left[ d\Omega_{S^2} ^2 + \left( \frac{d \zeta }{N} + \cos \theta d\phi \right) ^2 \right]+{\cal O}(R).    \label{metric0h}
\end{eqnarray}
This metric well behaves at the null surface $R=0$. The Killing vector field $V =\partial_v$ becomes null at $R=0$ and $V$ is hypersurface orthogonal from $V_\mu dx^\mu = g_{vR} dR$ at the place. 
Therefore the hypersurface $R = 0$ is a Killing horizon.  
In the coordinate system $(v ,R ,\theta,\phi,\zeta)$, 
each component of the metric is analytic in the region of $R\ge 0$. Hence the   space-time has no curvature singularity on and outside the black hole horizon.

The induced metric on the three-dimensional 
spatial cross section of the black hole horizon located at $R=0$ with the timeslice is obtained as  
\begin{eqnarray}
 \left. ds^2 \right| _{R=0,v={\rm const.}} 
 =\frac{LMn}{2} \left[ d\Omega _{S^2} ^2 + \left( \frac{d\psi}{n} + \cos \theta d\phi \right) ^2 \right] 
 = 2 L M n d\Omega _{\rm S^3/{\mathbb Z}_n} ^2,       
\end{eqnarray}
where $d\Omega _{\rm S^3/{\mathbb Z}_n} ^2$ denotes the metric on the lens space $L(n;1)={\rm S^3}/{\mathbb Z}_n$ with a unit radius. In particular, in the case of $n=1$, the shape of the horizon is a round $\rm S^3$ in contrast to the Gaiotto {\it et al}'s supersymmetric black holes~\cite{GSY}.

\subsection{Ergoregions}
Here, we investigate the number and the structure of the ergoregions. 
As is discussed in the ref.~\cite{TIMN}, the space-time admits a considerable 
rich structure by two kinds of rotations of black holes and the background.

For the solution \eqref{metric0}, the ergosurfaces are located at $R$ satisfying the equation, $f(R): = \left( 1-\alpha^2 \right) \left( R+M \right)^2 \left( R+N \right)^2g_{\bar t \bar t}=0$, where the explicit form of the function $f(R)$ is given by 
\begin{eqnarray}
f(R) &=& 
- \left( 1 - \alpha ^2 \right)
   R^4+\left[ 3 \left( M+N \right) \alpha
   ^2-2 N\right] R^3 \notag \\ 
& & +\left[ 3 M \alpha ^2 \left( M+N \right)
 -N^2 \left( 1-\alpha^2 \right)^2 \right]
   R^2+M^2 \left( M+3 N \right) \alpha ^2
   R+M^3 N \alpha ^2.              \label{defergor}
\end{eqnarray}

\begin{figure}[htbp]
 \begin{center}
 \includegraphics[width=10cm,clip]{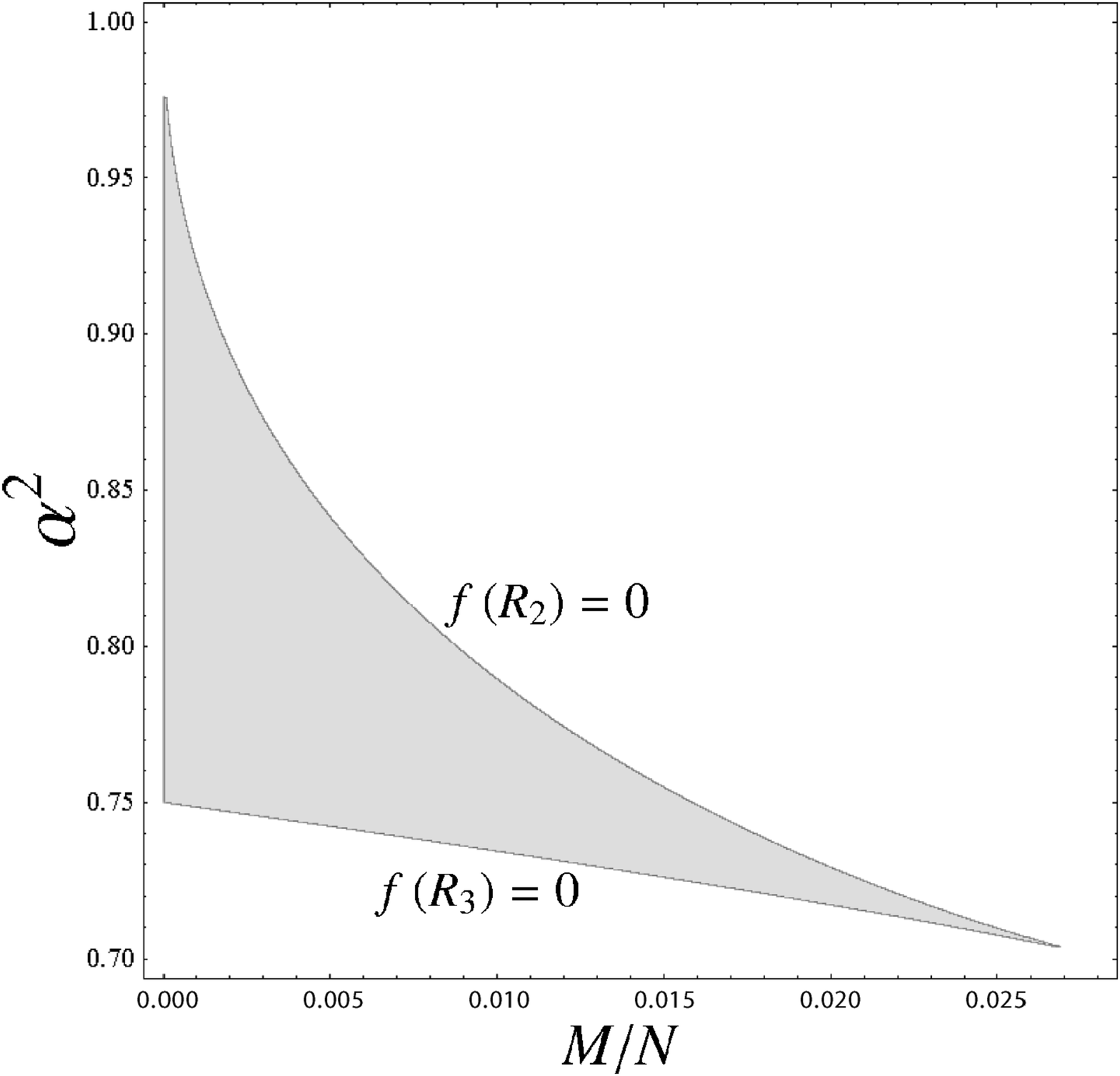}
 \end{center}
 \caption{This figure shows a region of parameters 
for the solution \eqref{metric0} with two disconnected ergoregions. }
 \label{2ergos}
\end{figure}

Note that $f(0) = M^3 N \alpha ^2 > 0$ and $f(\infty) < 0$ for the regions of parameters \eqref{parahanni}. Hence there always exists an ergoregion around the black hole horizon.  
Furthermore, 
within the parameter region $(M ,~N ,~\alpha )$ satisfying the inequalities, 
$f(R_1)f(R_2)f(R_3)<0$ and $f'(R_+)f'(R_-)<0$, which is shown 
in FIG.\ref{2ergos}, there are two disconnected ergoregions,
 inner  ($0 \le R \le R_{II}$) and outer $(R_{III} \le R \le R_{IV})$ ergoregions, 
where $R_i (i=1,~2,~3, ~R_1 < R_2 < R_3)$, $R_\pm (R_- < R_+)$ and 
$R_\gamma  (\gamma =I, \cdots, IV, ~ R_{I} < R_{II} < R_{III} < R_{IV})$ are three different positive roots of 
$f'(R)=0$, two different positive roots of $f''(R)=0$ and four different roots of $f(R)=0$, respectively. 
The inner ergoregion is inside a sphere which contains the black hole horizon. 
The outer ergoregion has a shape of shell, {\sl ergoshell}, which is disjoint 
from the inner ergoregion. 
There exists a normal region between the inner and the outer ergoregions.

The angular velocities of the locally non-rotating observers are obtained as
\begin{eqnarray} 
\Omega _{\phi} = 0 , \quad
\Omega _{\tilde \psi} = - 2\frac{\alpha ^2 N R^3 + M (R+N) \left( M^2 + 3 M R + 3 R^2 \right) }
{L \sqrt{1-\alpha ^2} \left[ (R+M)^3 (R+N) - \alpha ^2 R^4 \right]} \alpha .     
\end{eqnarray} 
From these equations, two disconnected ergoregions of the solution \eqref{metric0} always rotate in 
the same direction in contrast to the squashed Kerr-G\"odel black hole solutions \cite{TIMN}, which two ergoregions can also rotate in the opposite directions.  The angular velocities of the horizon $R=0$ are 
\begin{eqnarray} 
\Omega _{\rm H \phi} = 0 , \quad
\Omega _{\rm H \tilde \psi} = - \frac{2 \alpha }{L \sqrt{1-\alpha ^2}}.     
\end{eqnarray}

\section{Two Rotating Black Holes}\label{sec:multi}

For simplicity, we restrict ourselves to the two-black holes case, i.e., 
$M_i=N_i=0\  (i\ge 3)$. Without loss of generality, we can put the locations of two point sources as $\bm R_1=(0,0,d)$ and $\bm R_2=(0,0,-d)$, where the constant $2d$ denotes the separation between two black holes. 
\subsection{metric}
In this case, the metric is given by
\begin{eqnarray}
ds^2 = - H^{-2} \left[ dt + \alpha V \left( d\zeta +\bm \omega \right) \right] ^2
        + H \left[ 
V^{-1} \left( dR^2 + R^2 d\Omega _{S^2} ^2 \right) 
                  + V \left( d\zeta +\bm \omega \right) ^2
\right] ,       \label{2metric} 
\end{eqnarray}
where the functions $H,~V^{-1}$ and the one-form $\bm \omega $ are  
\begin{eqnarray}
  H &=& 1
+ \frac{M_1}{\sqrt{R^2-2d R\cos\theta+d^2}} 
+ \frac{M_2}{\sqrt{R^2+2d R\cos\theta+d^2}} , \\
 V^{-1} &=& 1 + \frac{N_1}{\sqrt{R^2-2d R\cos\theta+d^2}} 
+ \frac{N_2}{\sqrt{R^2+2d R\cos\theta+d^2}} , \\ 
 \bm \omega &=& \left( N_1 \frac{R\cos\theta - d  }{\sqrt{R^2-2d R\cos\theta+d^2}}
            + N_2 \frac{R\cos\theta + d }{\sqrt{R^2+2d R\cos\theta+d^2}} \right) d\phi .
\end{eqnarray}

The constants $N_i$ $(i = 1,~2)$ are related to the size of the compactified radius $L$ 
at the infinity by 
\begin{eqnarray}
N_i = \frac{L}{2} n_i,
\end{eqnarray}
where $n_i$ $(i = 1,~2)$ are the natural numbers.

\subsection{Near horizon}
The metric diverges at the locations of two point sources, i.e., $\bm R=\bm R_1$ and $\bm R=\bm R_2$. We make the coordinate transformation so that $\bm R_1=0$ and $\bm R_2=(0,0,-2d)$. Then, the functions $H,V^{-1}$ and the one-form $\bm \omega$ in the metric and the gauge potential take the following forms
\begin{eqnarray}
H &=& 1
+ \frac{M_1}{R} 
+ \frac{M_2}{\sqrt{R^2+4d R\cos\theta+4d^2}} , \\
 V^{-1} &=& 1 + \frac{N_1}{R} 
+ \frac{N_2}{\sqrt{R^2+4d R\cos\theta+4d^2}} , \\ 
 \bm \omega &=&  \left( N_1\cos\theta
            + N_2 \frac{R\cos\theta + 2d }{\sqrt{R^2+4d R\cos\theta+4d^2}} \right) d\phi ,
\end{eqnarray}
respectively.

In order to remove this apparent divergence at $\bm R=\bm R_1=0$, 
we introduce new coordinates $(v, \zeta' )$ such that 
\begin{eqnarray} 
 dv &=& dt - \left[ \frac{\left(3
   \left(2d+M_2\right) N_1+M_1
   \left(2d+N_2\right)\right)
   \sqrt{M_1}}{4d \sqrt{
   N_1} R}+\frac{\sqrt{M_1^3
   N_1}}{R^2} \right] dR , \\
 d\zeta ' &=& d\zeta + N_2 d \phi .   
\end{eqnarray}
Then, near $R = 0$, the metric \eqref{2metric} behaves as 
\begin{eqnarray} 
 ds^2 & \simeq & -  2\sqrt{\frac{N_1}{M_1}} dv dR+ M_1 N_1 
\left[\left(\frac{d\zeta '}{N_1} + \cos\theta d\phi\right)^2 + d\Omega_{\rm S^2} ^2 \right] 
\notag \\
& & + \left[ \frac{3 N_1 (2d+M_2) (N_1 (2d+M_2) + 2 M_1 (2d+N_2)) - M_1 ^2 (2d+N_2) ^2}
{16 d^2 M_1 N_1} \right. \notag \\
& & \left. - \frac{3 M_2 N_1 + M_1 N_2}{4d^2} \cos \theta \right] dR^2 . 
\end{eqnarray}

This metric well behaves at the null surface $R=0$. 
In this case, $g_{vv}=0$ and $(\partial_v )_\mu dx^\mu = g_{vR} dR$ also hold at $R=0$.  
Therefore, the Killing vector field $\partial_v$ becomes null and is hypersurface orthogonal at this place. 
So the  hypersurface $R = 0$ is a Killing horizon. From the same discussion, the other point source ${\bm R}=\bm R_2$  also corresponds to a Killing horizon.

Note that $\partial_\zeta=\partial_{\zeta'}$. So the periodic coordinate $\zeta'$ has the same periodicity as $\zeta$. Then, the induced metric on the $i$-th horizon $(i=1,2)$ is 
\begin{eqnarray}
	ds^2|_{i{\rm -th} \ {\rm horizon}}
	=\frac{LM_i n_i}{2}\left[\left(
		\frac{d\psi'}{n_i}+\cos\theta d\phi\right)^2 
		+d\Omega_{\rm S^2} ^2 \right] , 
\end{eqnarray} 
where $0 \leq \psi' = 2 \zeta' / L\leq 4\pi$. 
Hence the horizon is topologically the lens space $L(n_i;1)=\rm S^2/{\mathbb Z}_{n_i}$.

\subsection{Asymptotic structure}

In the neighborhood of the infinity, $R=\infty$, 
the harmonic functions $H$ and $V^{-1}$ behaves as ones with a single point source, i.e.,
\begin{eqnarray}
H &\simeq& 1+\frac{\sum_i M_i}{R}+{\cal O}\left(\frac{1}{R^2}\right),\\
V^{-1} &\simeq& 1+\frac{\sum_i N_i}{R}+{\cal O}\left(\frac{1}{R^2}\right).
\end{eqnarray}
Then, the one-form $\bm \omega$ is asymptotically
\begin{eqnarray}
{\bm \omega}\simeq \left(\sum_i N_i\right)\cos\theta d\phi+{\cal O}\left(\frac{1}{R}\right) .
\end{eqnarray}

Hence the metric asymptotically behaves as    
\begin{eqnarray}
 ds^2 \simeq -d\tilde t ^2 + dR ^2 + R ^2 d\Omega _{S^2} ^2 
  + \frac{L^2}{4} \left( 1-\alpha ^2 \right) \left( \frac{d\bar \psi}{\sum_in_i} + \cos \theta d\phi \right) ^2,    
\end{eqnarray}
where we introduced the following coordinates $(\tilde t,\bar \psi)$
\begin{eqnarray} 
\tilde t = \frac{t}{ \sqrt{1-\alpha ^2} } , \quad 
\bar \psi = \psi - \frac{\alpha t}{\sum_i N_i \left( 1-\alpha ^2 \right) },  
\end{eqnarray}
which are chosen so that they are in the rest frame at the infinity. 
The asymptotic structure of the solution \eqref{2metric} is an asymptotically locally flat, 
i.e., the metric asymptotes to
a twisted constant $S^1$ fiber bundle over the four-dimensional Minkowski   space-time and the spatial infinity has the structure of an $\rm S^1$ bundle over an $\rm S^2$ such that it is the lens space $L(n;1)=\rm S^3/{\mathbb Z}_n$, where $n=\sum_in_i$ is the natural number.

From the asymptotic behavior of the metric, we can obtain 
the Komar mass, the charge and the Komar angular momenta at the spatial infinity as
\begin{eqnarray}
M &=&
\frac{3 L \left[ \left( \alpha ^2 + 2 \right) \sum _i M_i + \alpha ^2
\sum _i N_i \right]}
{ 8 \pi \sqrt{ 1-\alpha ^2 } G_5 }\frac{{\cal A}_{\rm S^3}}{ n}, \\
Q &=& \frac{\sqrt 3 L \sum _i M_i }
{ 2 \pi G_5 }\frac{{\cal A}_{\rm S^3}}{ n} , \\
J_{\phi} &=& 0 , \\
J_{\tilde \psi} &=& - \frac{\alpha L ^2 \left( 3 \sum _i M_i + \alpha ^2
\sum _i N_i \right)}
{ 8 \pi G_5 } \frac{{\cal A}_{\rm S^3}}{n}.
\end{eqnarray}

\subsection{Ergo regions}

For simplicity, 
assume that two black holes have the equal mass and 
the horizon topology of $\rm S^3$, which correspond to the choice of the parameters $M_1=M_2$ and $N_1=N_2=L/2$.
The ergosurfaces are located at $R$ satisfying the equation
\begin{eqnarray}
-H^{-2}\left(\sqrt{1-\alpha^2}+\frac{\alpha^2}{\sqrt{1-\alpha^2}}V\right)^2+\frac{\alpha^2}{1-\alpha^2}H V=0.\label{eq:ergo_2}
\end{eqnarray}

\begin{figure}[htbp]
  \begin{center}
   \includegraphics[width=12.5cm,clip]{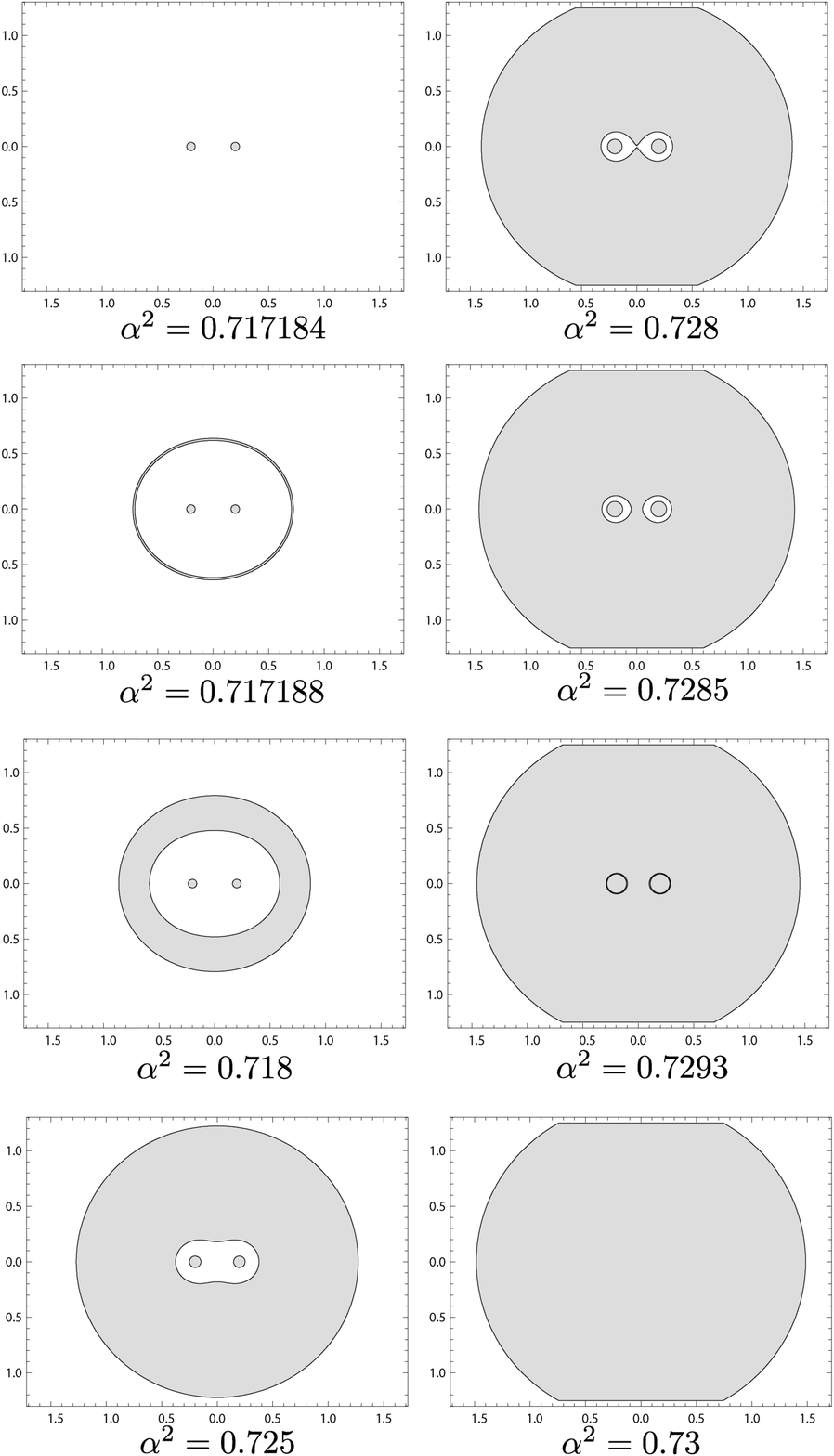}
  \end{center} 
  \caption{Ergoregions (shaded regions) in varying $\alpha^2$. 
	$d=0.2$ and $\alpha^2=0.717184 \sim 0.73$. }
\label{fig:rot1}
\end{figure}

\begin{figure}[htbp]
  \begin{center}
   \includegraphics[width=12.5cm,clip]{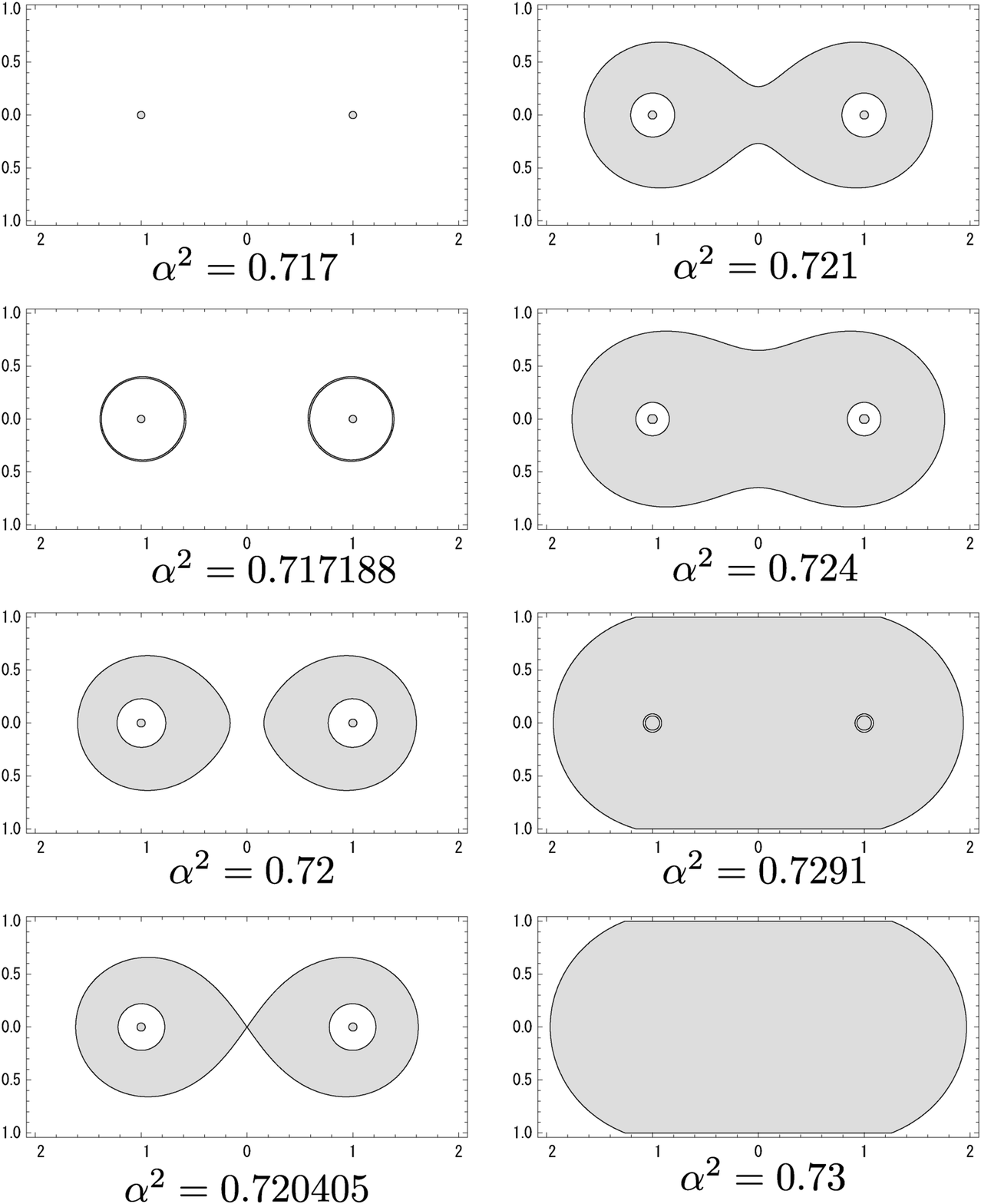}
  \end{center} 
  \caption{Ergoregions in varying $\alpha^2$. 
	$d=1$ and $\alpha^2=0.717 \sim 0.73$. }
\label{fig:rot2}  
\end{figure}

Introduce the coordinates $(x,y)$ defined by $x=R\cos\theta$ and 
$y=R\sin\theta$. 
FIG.\ref{fig:rot1} and FIG.\ref{fig:rot2} show how the ergoregions change the shapes in a $(x,y)$-plane as the rotation parameter $\alpha$ varies with 
the other parameters fixed in the cases of the separation parameter 
$d=0.2$ and $d=1$, respectively.  
Here the horizontal axis and the vertical axis denote the $x$-axis 
and the $y$-axis, respectively. In these figures, the shaded regions denote 
the ergoregions, i.e., the regions such that the left hand side of 
Eq.(\ref{eq:ergo_2}) is positive. Two black holes are located at $(\pm d,0)$ 
in this plane.

First, see FIG.\ref{fig:rot1}, the case of $d=0.2$. 
There exists an ergoregion around each rotating black hole when $\alpha$ is small.
When $\alpha^2\simeq 0.717188$, a new ergoregion in the shape of a shell 
enclosing two black holes appears far away from them. 
 There is an inner normal region between the ergoshell 
and the two inner ergoregions. 
As the value of $\alpha$ gets larger, the outer ergoshell becomes thick. 
When $\alpha^2\simeq 0.728$, the inner normal region becomes 
disconnected.
When $\alpha^2\simeq 0.7293$,  the outer ergoshell and the two inner   
ergoregions merge and the inner normal region disappears.

See also FIG.\ref{fig:rot2}, the case of $d=1$. 
In this case, when $\alpha^2\simeq 0.717188$, in contrast to the previous case,
two disconnected outer ergoshells appear around each black hole. 
When $\alpha^2 \simeq 0.720405$ they merge together.

\begin{figure}[htbp]
  \begin{center}
   \includegraphics[width=12.5cm,clip]{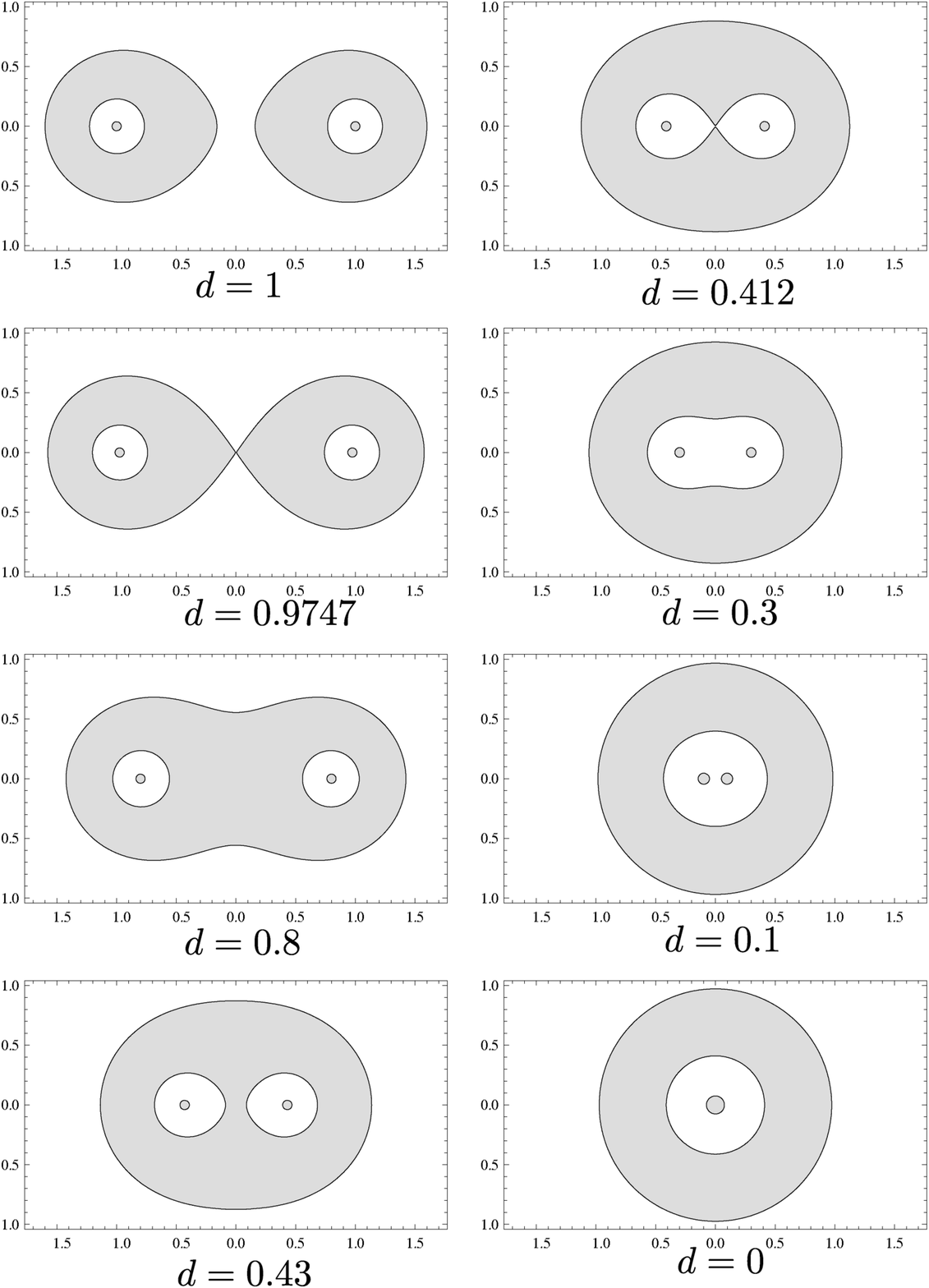}
  \end{center} 
 \caption{Ergoregions in varying $d$. $d = 1 - 0$ and $\alpha ^2 = 0.72$.  }
\label{fig:sep}  
\end{figure}

Next, 
FIG.\ref{fig:sep} shows how the shapes of ergoregions change with varying 
the separation parameter $d$, where the other parameters are kept unchanged. 
When the separation is large enough, there exist two disconnected outer ergoshells 
and two inner ergoregions around each black hole. 
When two black holes become closer, two outer ergoshells are connected 
with each other into a single large outer ergoshell. 
There are two inner normal regions around each black hole. 
When the separation becomes smaller, the inner normal regions join together. 
Finally, two inner ergoregions around two black holes also join together.

\section{Summary and Discussion}\label{sec:summary}

We have considered the limiting case given by (\ref{limit}) in the 
Kaluza-Klein-Kerr-Newman-G\"odel black hole solutions~\cite{TIMN}, 
and have presented supersymmetric 
Kaluza-Klein-Kerr-Newman-G\"odel multi-black hole solutions,  
in the five-dimensional Einstein-Maxwell theory with a Chern-Simons term. 
The new solutions have no closed timelike curve everywhere outside the 
black hole horizons. 
At the infinity, the metric asymptotically approaches a twisted 
$\rm S^1$ bundle over the four-dimensional Minkowski space-time.

Though the Kerr parameter and the G\"odel parameter are related, 
each black hole can have an inner ergoregion and an outer ergoshell depending on 
the parameter. We have explicitly presented the various shapes of 
ergoregions in the case of two-black holes.

The solutions (\ref{metric})-(\ref{omega}) can be easily generalized to solutions with a positive cosmological constant $\Lambda>0$. In this solutions, the harmonic function (\ref{H}) is replaced by
\begin{eqnarray}             
 H &=& \lambda t + \sum _i \frac{M_i}{\left| \bm R - \bm R _i \right|} ,\label{eq:cos}
\end{eqnarray}
where the constant $\lambda$ is related to the cosmological constant by $\lambda = \pm 2 \sqrt{\Lambda / 3}$. In particular, in the case with $\beta=-1$ and $\epsilon=0$, the solutions coincide with five-dimensional Kastor-Traschen solutions $(\alpha = 0)$~\cite{London} or Klemm-Sabra solutions $(\alpha \neq 0)$~\cite{KS}, which describes the coalescences of black holes with the horizon topologies of $\rm S^3$ 
into a single black hole with the horizon topology of $\rm S^3$.  
In the case of two black holes with $M_i \neq 0,~N_i \neq 0 ~ (i=1,~2)$, 
$M_j =N_j = 0 ~ (j \geq 3)$, $\beta=-1$ and $\epsilon=0$, the solution
describes the coalescence of two rotating black holes 
with the horizon topologies of $\rm S^3$  
into a single rotating black hole 
with the horizon topology of the lens space $L(2;1) = {\rm S^3} / {\mathbb Z} _2$ \cite{MIKT}. Cosmological non-rotating multi-black hole solutions on the multi-centered-Taub-NUT space are obtained by setting $\alpha = 0,~\beta = -1 $ and $\epsilon = 1$ \cite{IIKMMT}. This solution is not static even in a single black hole case. 
In the case of $\beta=1$, the solution with the harmonic function (\ref{eq:cos}) would also describe 
the coalescence of black holes. We leave the analysis for the future.

\section*{Acknowledgments} 
H.I. is supported by the Grant-in-Aid for Scientific Research No. 19540305.  
S.T. is supported by the JSPS under Contract No. 20-10616.

\appendix

\section{Parameter Region}\label{PR}

Here we show the inequality $0\le \alpha^2<1$ under $M_i>0$ and $N_i>0$ is the necessary and sufficient condition for the absence of closed timelike curves outside the horizons.

Assume that all point sources are located at the $z$-axis,  i.e., $\theta=0,\pi$ on the three-dimensional Euclid space in the Gibbons-Hawking space. In this case, the one-form $\bm\omega$ is proportional to $d\phi$.
The condition of the absence of closed timelike curves outside the horizons is equivalent to the condition that the two-dimensional $(\phi,\zeta)$-part of the metric, 
\begin{eqnarray}
ds^2|_{(\phi,\zeta)} &=& A(d\zeta+\bm\omega)^2+Bd\phi^2
\end{eqnarray}
is positive-definite, where 
\begin{eqnarray}
A&=&HV-\alpha^2H^{-2}V^2,\\
B&=&H V^{-1}R^2\sin^2\theta.
\end{eqnarray}

This metric is positive-definite if and only if the following two-dimensional matrix is positive-definite
\begin{eqnarray}
M=\left(
\begin{array}{cc}
A& 0\\
0& B
\end{array}
\right) .
\end{eqnarray}
Therefore, noting that $B>0$, we obtain the condition
\begin{eqnarray}
M > 0 \Longleftrightarrow A > 0 .
\end{eqnarray}
As a result, it is enough to prove the $A >0$. At the infinity $R \to \infty$, the function behaves as 
\begin{eqnarray}
A\simeq 1-\alpha^2+{\cal O}\left(\frac{1}{R}\right).
\end{eqnarray}
Hence $0\le \alpha^2<1$ is necessary. Next we show that if $0\le \alpha^2<1$  is satisfied with $M_i>0$ and $N_i>0$, there is no closed timelike curve outside the horizons.
Noting that $1<H<\infty$ and $1<V^{-1}<\infty$ under the conditions $M_i>0$ and $N_i>0$, the inequality
\begin{eqnarray}
A& = & H V(1-\alpha^2H^{-3}V) \notag \\
 & > & H V(1-\alpha^2) \notag \\
 & > & 0
\end{eqnarray}
holds everywhere outside the horizons.

\section{Squashed Kerr-Newman-G\"odel black holes}\label{KN}
The metric and the gauge potential of the squashed Kerr-Newman-G\"odel black hole solution \cite{TIMN}  
is given by
\begin{eqnarray}
ds^2 &=& -f(r) dT ^2-2g(r) dT \left( d \psi + \cos \theta d\phi \right) + 
h(r) \left( d \psi + \cos \theta d\phi \right) ^2 \notag \\
& &+ \frac{k^2(r)}{V(r)} dr^2 
+ \frac{r^2}{4}[k(r) d\Omega_{S^2}^2 + \left( d \psi + \cos \theta d\phi \right) ^2],
\end{eqnarray}
and
\begin{eqnarray}
{\bm A}=\frac{\sqrt{3}}{2}
\left[\frac{q}{r^2}dT
+\left(jr^2+2jq-\frac{qa}{2r^2}\right)
\left( d \psi + \cos \theta d\phi \right) 
\right],
\end{eqnarray}
respectively, where the metric functions are 
\begin{eqnarray}
f(r)&=&1-\frac{2m}{r^2}+\frac{q^2}{r^4},\\
g(r)&=&jr^2+3jq+\frac{(2m-q)a}{2r^2}-\frac{q^2a}{2r^4},\\
h(r)&=&-j^2r^2(r^2+2m+6q)+3jqa+\frac{(m-q)a^2}{2r^2}-\frac{q^2a^2}{4r^4},\\
V(r)&=&1-\frac{2m}{r^2}+\frac{8j(m+q)[a+2j(m+2q)]}{r^2}\nonumber\\
    & &+\frac{2(m-q)a^2+q^2(1-16ja-8j^2(m+3q))}{r^4},\\
k(r)&=&\frac{V(r_\infty)r_\infty^4}{(r^2-r_\infty^2)^2}.
\end{eqnarray}
In the limit of $r_\infty \to \infty$, i.e., $k(r)\to 1$, the solution coincides with the Kerr-Newman-G\"odel black hole solution in Ref.\cite{WU}.

\end{document}